\documentclass[a4paper,11pt]{article}
\usepackage{pos}
\usepackage{comment}
\newcommand \mres {m_{\mathrm{res}}}
\newcommand \barpsi {\langle\bar{\psi}\psi\rangle}

\title{Three flavor QCD phase transition with M\"{o}bius domain wall fermions}

\author*[a]{Yu Zhang}
\author[b]{Yasumichi Aoki}
\author[c,d]{Shoji Hashimoto}
\author[b]{Issaku Kanamori}
\author[c,d,e]{Takashi Kaneko}
\author[b]{Yoshifumi Nakamura}

\affiliation[a]{Fakult\"{a}t f\"{u}r Physik, Universit\"{a}t Bielefeld, D-33615 Bielefeld, Germany}

\affiliation[b]{RIKEN Center for Computational Science,  7-1-26
	\\	Minatojima-minami-machi, Chuo-ku, Kobe, Hyogo 650-0047, Japan}

\affiliation[c]{High Energy Accelerator Research Organization (KEK), Tsukuba 305-0801, Japan}

\affiliation[d]{School of High Energy Accelerator Science, The Graduate University for Advanced Studies (Sokendai), Tsukuba 305-0801, Japan}

\affiliation[e]{Kobayashi-Maskawa Institute for the Origin of Particles and the Universe, Nagoya University, Nagoya 464–8602, Japan}

\emailAdd{yzhang@physik.uni-bielefeld.de}

\abstract{We present an updated study of the $N_f=3$ QCD phase transition using M\"{o}bius domain wall fermions. Simulations were performed on $N_t=12$ lattices with aspect ratios ranging from 2 to 4 for various quark masses, at a lattice spacing of $a=0.1361(20)$ fm, corresponding to a temperature of 121(2) MeV. To clarify the nature of the phase transition, a large-volume lattice, $48^3 \times 12\times 16$, was added to analyze the volume dependence of disconnected chiral susceptibility. By examining  the chiral condensate, disconnected chiral susceptibility, and Binder cumulant, and incorporating results from $24^3 \times 12 \times 16$ and $36^3 \times 12 \times 16$ lattices reported in earlier studies~\cite{Zhang:2022kzb,Zhang:2024ldl}, we 
	observe that the transtion is consistent with a crossover
	at a quark mass of approximately $m_f^{\mathrm{\overline {MS}}}(2\, \mathrm{GeV}) \sim 4$ MeV at this temperature. Furthermore, we discuss the effects of residual chiral symmetry breaking on the chiral condensate and disconnected chiral susceptibility for different sizes in the 5th-direction.}

\FullConference{The 41st International Symposium on Lattice Field Theory (LATTICE2024)\\
28 July - 3 August, 2024\\
Liverpool, UK\\}

\begin{document}
\maketitle

\section{Introduction}
The nature of the QCD chiral phase transition has been a hot topic in the study of strongly interacting matter for decades. In particular, the phase transition for three degenerate quark flavors in the chiral limit, represented in the lower left corner of the Columbia plot, remains unresolved. Early theoretical work by Pisarski and Wilczek, based on a perturbative renormalization group study of a three-dimensional linear sigma model, predicted a first-order phase transition~\cite
{Pisarski:1983ms}. According to their analysis, this first-order transition weakens as one moves away from the chiral limit and eventually terminates at a critical endpoint, where the phase transition becomes second-order, belonging to the three-dimensional $Z(2)$ Ising universality class.

In contrast, recent functional renormalization group analysis with all couplings up to $\phi^6$ terms included in a three-dimensional Ginzburg-Landau theory~\cite
{Fejos:2022mso,Fejos:2024bgl}, numerical conformal bootstrap~\cite{Kousvos:2022ewl} 
, conjectures in Ref.~\cite{Pisarski:2024esv} and extended linear sigma model analysis~\cite{Giacosa:2024orp}, both suggest the possibility of a second-order phase transition in the $N_f = 3$ chiral limit if $U(1)_A$ symmetry is effectively restored at the critical point. Recent results from the Dyson-Schwinger equations~\cite{Bernhardt:2023hpr} also found a second-order phase transition in the $N_f = 3$ chiral limit. These low-energy effective model analyses can capture qualitative aspects but certainly cannot solve this open problem. Whether the first-order region is tiny or does not exist at all is still an open question. If it exists, what is the value of the critical pion mass that separates the first-order and crossover regions? To answer this quantitative question, we need lattice QCD simulation.

Earlier lattice studies using standard staggered~\cite{Liao:2001en,Karsch:2001nf,deForcrand:2003vyj}, Wilson~\cite{Iwasaki:1995ij} and $O(a)$-improved Wilson fermions~\cite
{Jin:2014hea} support the existence of a first-order phase transition for light quark masses on coarse lattices. However, subsequent studies have shown that the size of the first-order region depends significantly on the formulation of the lattice action and lattice spacing~\cite{Karsch:2003va,deForcrand:2006pv,Varnhorst:2015lea,Bazavov:2017xul,Jin:2017jjp,Kuramashi:2020meg}. 
Recent studies have not found direct evidence of a first-order transition for pion mass above 50 MeV using highly improved staggered fermions(HISQ)~\cite{Bazavov:2017xul} or 110 MeV using $O(a)$-improved Wilson fermions~\cite{Kuramashi:2020meg}. 
 Another recent study by looking into the position of the tri-critical point as a function of the number of quark flavors, $N_f$,
finds that the chiral phase transition is second order for three massless quarks~\cite{Cuteri:2021ikv}. Later, a study using HISQ fermions also suggests a second-order phase transition in the chiral limit~\cite
{Dini:2021hug}. This leaves little room for a first-order region to exist, but it cannot be ruled out; it can be very small. It is important to note that the existing results predominantly originate from staggered or Wilson fermion formulations, which partially or completely break the chiral symmetry at finite lattice spacing. Therefore, it is important to explore this problem further using chiral fermions. Specifically, we use the M\"{o}bius domain wall fermion, and some of the preliminary results have been reported in previous lattice conference proceedings~\cite{Nakamura:2022abk,Zhang:2022kzb,Zhang:2024ldl}. The advantage of the M\"{o}bius domain wall fermion is that it has exact chiral symmetry at finite lattice spacing for the size of fifth-dimension ($L_s$) goes to infinity, and the reduced chiral symmetry breaking effect parameterized by the residual mass 
 when the size in the 5th-direction, $L_s$, is finite.

 \section{Lattice Setup}
We perform $N_f=3$ QCD simulations using the tree-level improved Symanzik gauge action combined with the Möbius domain wall fermion  action. The simulations are implemented using the Grid code optimized for the Fugaku CPU A64FX architecture~\cite{Meyer:2019gbz}. The gauge coupling is fixed at $\beta=4.0$, corresponding to a lattice spacing $a=0.1361(20)$ fm. This value is obtained from the Wilson flow parameter $t_0$ and matches the continuum extrapolated physical point result for $N_f=2+1$ QCD~\cite{Borsanyi:2012zs}. In our previous study~\cite{Zhang:2022kzb,Zhang:2024ldl}, we performed simulations on $24^3\times12\times16$ and $36^3\times12\times16$ lattices with a range of quark masses. We observed a significant finite-volume effect near the transition point. To study this effect further, in this work, we conducted a larger volume simulation on $48^3\times12\times16$ lattices with two mass points $-$0.003 and $-$0.004. To evaluate the effects of residual chiral symmetry breaking, additional simulations are performed on $24^3\times12\times32$ lattices with 5 quark masses in the range $\left[ -0.001, 0.003\right]$. For each parameter set, we generate about 20,000 thermalized trajectories, with measurements taken every 10 trajectories. 

In addition to the finite-temperature ensembles, we also generate zero-temperature $N_f=3$ configurations at $\beta=4.0$, 4.1 and 4.17 on lattices of size $24^3\times48\times16$, $24^3\times48\times16$, and $32^3\times64\times16$, respectively. These ensembles are utilized to determine the lattice spacing, study the coupling dependence of the residual mass, and remove the ultraviolet divergence contribution from the finite temperature chiral condensate.

\section{Numerical results}

\subsection{Residual chiral violations}
For domain wall fermions with finite $L_s$, there exists a residual mixing between the two walls, resulting in a breaking of chiral symmetry. The leading effect of this mixing is an additive renormalization to the bare quark mass, commonly referred to as the residual mass, $m_{\rm{res}}$. 
The residual mass is defined through the Ward-Takahashi identity of the axial current~\cite{Furman:1994ky}:
\begin{equation}
	\label{eq:WTI}
\Delta_{\mu}\langle \mathcal{A}^a_{\mu}(x) O(y) \rangle = 2m_f\langle J_5^a(x)O(y)\rangle + 2 \langle  J^a_{5q}(x)O(y)\rangle +  i\langle \delta^aO(y)\rangle 
\end{equation}
where $J^a_{5q}$ is the pseudoscalar density constructed from the fields at the center of the fifth dimension, and $J^a_{5}$ is the physical pseudoscalar density constructed from the fields at the boundary. The $J^a_{5q}$ term acts as an additional contribution to the continuum expression and vanishes in the limit $L_s \to \infty$ for the non-singlet flavor. Close to the continuum limit, in an effective low-energy Lagrangian, the coefficient of the mass term is expected to be proportional to $m_f + m_{\rm{res}}$. So, $J^a_{5q} \approx m_{\rm{res}} J^a_5$. This motivates the definition of the residual mass as the ratio~\cite{Blum:2000kn}
\begin{equation}
  \label{eq:ratio}
 m_{\rm{res}} = \frac{\left\langle \sum_{\vec x}J_{5q}^{a}(\vec{x},t)\,J_5^{a}(\vec{0},0) \right\rangle}{\left\langle \sum_{\vec x}J_5^a(\vec{x},t)\,J_5^a(\vec{0},0)\right\rangle} \Bigg|_{t \ge t_{min}} \,,
\end{equation}
where, for $t$ larger than some source-sink separations $t_{min}$, the ratio develops a plateau and equal to the residual mass. According to chiral perturbation theory, at leading order, pion mass squared is proportional to  $m_f + m_{\rm{res}}$. This dependence ensures that the pion mass vanishes in the chiral limit. 

\begin{figure}[!htp]
	\centering
     \includegraphics[width=0.42\textwidth]{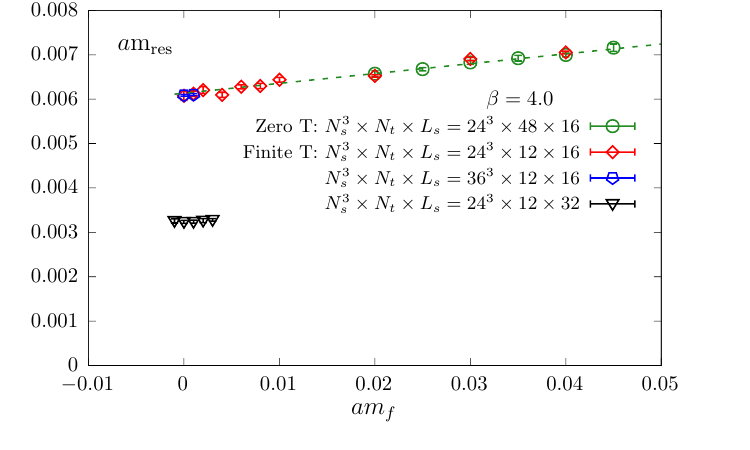}
\includegraphics[width=0.42\textwidth]{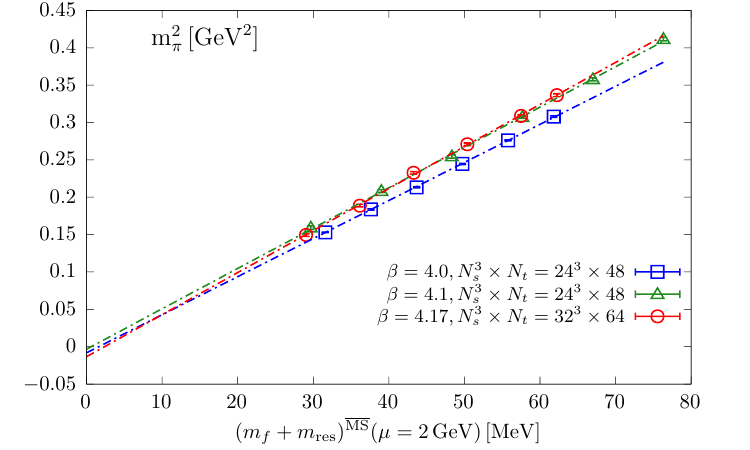}
\caption{Left: Residual mass as a function of the bare input quark mass $am_q$ for zero and finite temperature ensembles at $\beta=4.0$, with linear $am_f \to 0$ extrapolation to determine the mass independent $m_{\rm{res}}$. Right: The pion mass squared as a function of the renormalized quark mass in physical units for three different $\beta$ values. The dashed lines represent linear fits.} 
  	\label{fig:mres_mpi}
\end{figure}
To quantify the violation of chiral symmetry due to the finite $L_s$, we measured $m_{\rm{res}}$. The left plot of~\autoref{fig:mres_mpi} shows $m_{\rm{res}}$ as a function of bare input quark mass for zero temperature and finite temperature lattices at $\beta=4.0$. For finite temperature lattices, $m_{\rm{res}}$ is obtained using the ratio of correlation function evaluated at spatial source sink separation, given the larger spatial extent. We observe that the $m_{\rm{res}}$ obtained from finite-temperature and zero-temperature lattices are consistent, and also no visible volume dependence.
The dashed line represents a linear fit, which describes the data well and indicates a linear quark mass dependence of $m_{\rm{res}}$. This linear dependence is understood as a lattice artifact. A common approach to address this artifact and define a mass independent $m_{\rm{res}}$ is to extrapolate to the zero input quark mass limit. The extrapolated values are $am_{\rm{res}}(am_f=0) = 0.00613(9)$ for $L_s=16$ 
and $am_{\rm{res}}(am_f=0) = 0.00324(3)$ for $L_s=32$. This $L_s$ dependence is expected, as at strong gauge-field coupling, $m_{\rm{res}}$ 
receives significant contributions from near-zero eigenmodes, which gives a power-law dependence of $1/L_s$~\cite{RBC:2008cmd}. The 
$1/L_s$ dependence arises due to the presence of gauge field dislocations. 
The right plot of~\autoref{fig:mres_mpi} shows the pion mass squared as a function of the renormalized quark mass in the $\overline{\mathrm {MS}}$ scheme at a scale of 2 GeV for three different $\beta$ values. The results exhibit good linearity, as indicated by the dashed lines. This behavior is consistent with the leading order prediction of chiral perturbation theory, where $m_{\pi}^2 \propto m_{f} + m_{\rm{res}}$. As expected, $m_{\pi}$ approaches zero in the chiral limit, demonstrating the good chiral properties of domain wall fermions. However, $m_{\pi}^2$ does not reach exactly zero, which could be attributed to the finite-volume effects or the omission of chiral logarithm terms.
\subsection{Chiral condensate}
\begin{figure}[!htp]
	\centering
  \includegraphics[width=0.42\textwidth]{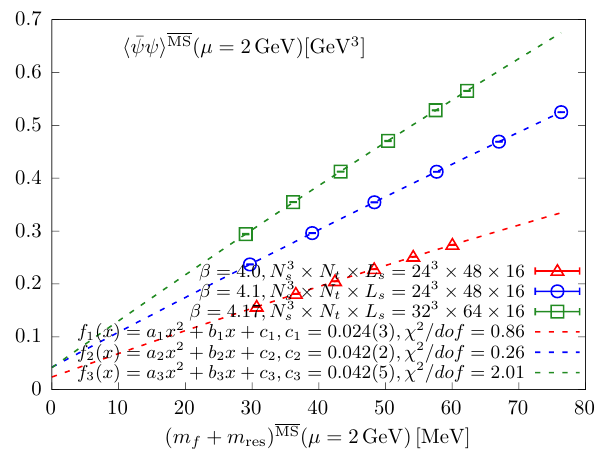} 
\includegraphics[width=0.42\textwidth]{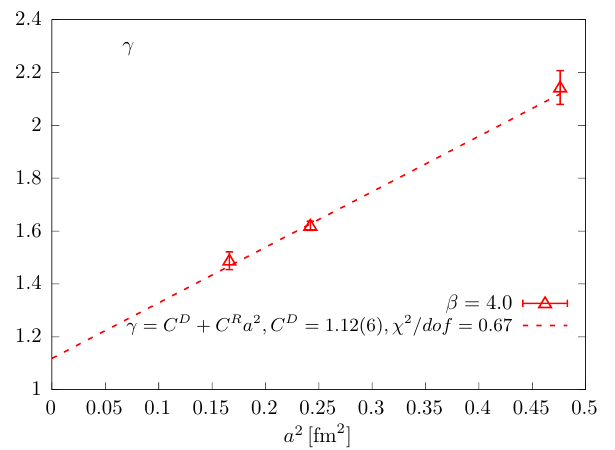}
  	\caption{Left: The multiplicatively renormalized chiral condensate as a function of the renormalized quark mass for three different $\beta$ values, with the dashed lines representing the quadratic fits. Right: The coefficient of linear quark mass term as a function of lattice spacings. }
  	\label{fig:zero_T_pbp}
\end{figure}

The chiral condensate serves as the order parameter of chiral symmetry breaking. To ensure that the chiral condensate remains finite in the continuum limit with a finite quark mass, both additive and multiplicative renormalizations are necessary. The additive renormalization arises from the $m_f a^{-2}$ divergent contribution. For domain wall fermion with finite $L_s$, there is an additional power divergence term induced by the residual chiral symmetry breaking, which is proportional to $xm_{\rm{res}}a^{-2}$ with $x$ an unknown coefficient~\cite
{Sharpe:2007}. Therefore, the domain wall fermion chiral condensate behaves as:
\begin{equation}\label{eq:pbp}
	\barpsi|_{\mathrm{DWF}} \sim \barpsi|_{\mathrm{cont.}} + C^D\frac{m_f + x\mres}{a^2} + ...\,,
\end{equation}
$m_f$ and $\mres$ do not appear in linear combination, in other words, $x=\mathcal{O}(1)$ instead of 1. If we perfrom extrapolation to the chiral limit $m=m_f + \mres \to 0$, the additive divergence remains, as shown as $\lim_{m\to 0} \lim_{L \to 0}\barpsi|_{\mathrm{DWF}} \sim \barpsi|_{\mathrm{cont.}} + C^D\frac{(x-1)\mres}{a^2}$.
The multiplicative renormalization is addressed using the renormalization constant $Z_m^{\overline{\rm{MS}}}(2\,\rm{GeV})$. In this work, the renormalized chiral condensate is defined as $(\barpsi - C^D \frac{m_f + x\mres}{a^2})(Z_m^{\overline{\mathrm {MS}}}(2\,\mathrm{GeV}))^{-1}$. 
To calculate it, 
the parameters $C^D$ and $x$ must be determined. Next, I will explain how we can get them. The left plot in~\autoref{fig:zero_T_pbp} shows the multiplicatively renormalized chiral condensate on zero temperature lattices for three different $\beta$ values, plotted as a function of the renormalized quark mass. The dashed lines represent quadratic fits, expressed as  follows: 
 \begin{align}\label{eq:chi_condensate_fit}
 	\langle \bar{\psi}\psi \rangle(m_f + m_{\rm {res}}) & = 	\langle \bar{\psi}\psi \rangle(0) + C^D \frac{m_f + x m_{\rm{res}}}{a^2} + C^R (m_f + m_{\rm{res}}) + A(m_f + m_{\rm{res}})^2 \\
 	&= 	\langle \bar{\psi}\psi \rangle(0) + (C^D + C^R a^2) \frac{m_f + m_{\rm {res}}}{a^2} + C^D \frac{(x-1)m_{\rm{res}}}{a^2} + A(m_f + m_{\rm{res}})^2\,.
 \end{align}
 The linear term in the quark mass receives contributions from both the UV divergence part and the regular part. These fits give a good descripation of the data. Right plot of~\autoref{fig:zero_T_pbp} shows the coefficient of the linear mass term as a function of lattice spacings. The value of $C^D$ is estimated by extrapolating to the continuum limit, as shown by the dashed line, resulting in $C^D=1.12(6)$. To eliminate the residual chiral symmetry breaking effect in the chiral condensate, it is necessary to calculate coefficient $x$, which can be obtained from the finite temperature chiral condensate.

\begin{figure}[!htp]
	\centering
\includegraphics[width=0.42\textwidth]{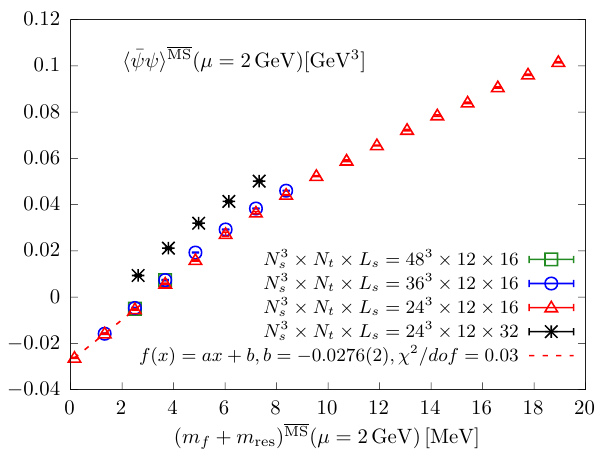}
\includegraphics[width=0.42\textwidth]{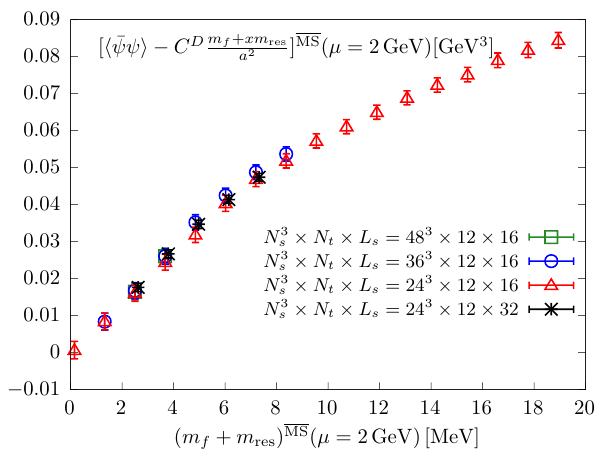}
  		\caption{The multiplicatively renormalized chiral condensate (left) and the additively and multiplicatively renormalized chiral condensate (right) as functions of the renormalized quark mass for $N_t=12$ lattices with $L_s=16$ and $L_s=32$ respectively.}
  	\label{fig:finite_t_pbp}
\end{figure}

The left panel of~\autoref{fig:finite_t_pbp} shows the multiplicatively renormalized chiral condensate as a function of the renormalized quark mass for finite tempearure ensembles with $N_t=12$, aspect ratios  $N_s/N_t$ from 2 to 4, and $L_s=16$ and $32$, respectively. We do not observe significant finite-volume effects, but we do see at approximately the same total quark mass, the chiral condensate for $L_s=32$ is always larger than the values obtained from $L_s=16$ lattices. This discrepancy arises from the residual chiral symmetry breaking term $C^D\frac{(x-1)\mres} {a^{2}}$, where it is negative and $m_{\rm{res}}$ is smaller for $L_s=32$. To remove this effect, a linear extrapolation to the chiral limit was performed using the three lowest mass points from the $24^3\times 12\times 16$ lattices. These mass points are believed to lie within the chiral symmetry restored phase, as the transition mass point determined from the peak of disconnected chiral susceptibility is above the chosen mass points at this temperature which we will show later and also the chiral phase transition temperature is estimated to be $T_c = 98^{+3}_{-6}$ MeV for $N_t=8$ from~\cite{Dini:2021hug}, so $\barpsi|_{\mathrm{cont}}$ is assumed to be zero. In the chiral limit,  the chiral condensate is expected to be $C^D\frac{(x-1)\mres} {a^{2}}$. By using the intercept value of the dashed line, along with the results for $C^D$ and $\mres$, $x$ is estimated to be $-0.6(1)$. This allows us to eliminate the intricate UV divergence from the chiral condensate. The right panel of~\autoref{fig:finite_t_pbp} shows the renormalized subtracted chiral condensate results. We observe that the results from $24^3\times 12\times16$ and $24^3\times 12\times 32$ lattices are consistent, indicating that the residual chiral symmetry breaking effect has been effectively removed from the chiral condensate. 
It is challenging to identify the transition point directly from
the chiral condensate itself. 
 Instead, we use the disconnected chiral susceptibility, which measures fluctuations in the chiral condensate and its response to changes in external parameters, such as the quark mass in this case.

\subsection{Disconnected chiral susceptibility}
\begin{figure}[!htp]
	\centering
\includegraphics[width=0.42\textwidth]{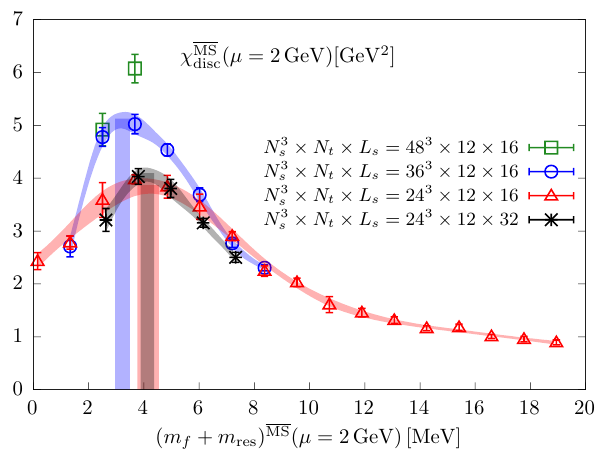}
\includegraphics[width=0.42\textwidth]{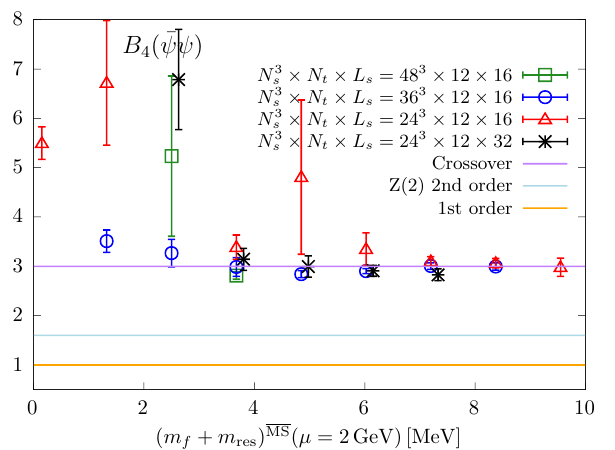}
    	\caption{Left: The renormalized disconnected chiral susceptibility as a function of the renormalized quark mass for $N_t=12$ lattices with $L_s=16$ and $L_s=32$, along with their cubic  spline fits, except for the largest volume. The vertical bands represent the transition region. Right: Binder cumulant as a function of the renormalized quark mass for $N_t=12$ lattices with $L_s=16$ and $L_s=32$.}
  	\label{fig:chidisc}
\end{figure}

We use the disconnected chiral susceptibility, $\chi_{\rm{disc}}$, to pinpoint the transition point where it develops a peak. $\chi_{\rm{disc}}$ does not suffer from the UV divergence, but multiplicative renormalization is required. 
The left panel of~\autoref{fig:chidisc} shows the result of the renormalized disconnected chiral susceptibility as a function of the renormalized quark mass in physical units for $L_s=32$ and $L_s=16$ lattices with three different volumes, along with their cubic smoothing spline fits, except for the largest volume. The vertical bands represent the transition region, determined from the peak of $\chi_{\rm{disc}}$. We observe a significant finite-volume effect for $N_{s}^3 \times 12 \times 16$ lattices with $N_{s} = 24, 32, 48$ near the transition range, but the change in peak height is not as large as anticipated from a first-order or $Z(2)$ second-order phase transition. This observation is consistent with a crossover rather than a true phase transition. Another observation is that at approximately the same total quark mass, $\chi_{\rm{disc}}$ obtained from $L_s=32$ lattices is consistent with those obtained from $L_s=16$ lattices with $N_s=24$ and $N_t=12$, despite the relative size of the residual mass and the input quark mass being quite different. This is expected, as the chiral susceptibility is defined as the derivative of chiral condensate with respect to the input quark mass, making it a function of the total quark mass. 

\subsection{Binder cumulant and Histogram of chiral condensate}
To identify the order of the QCD phase transition, we use the Binder cumulant of the chiral condensate~\cite
{Binder:1981sa}, which is defined as $B_4(\bar\psi \psi) = \frac{\left\langle (\delta \bar\psi \psi)^4\right\rangle}{\left\langle (\delta\bar\psi\psi)^2\right\rangle^2},\quad \delta\bar\psi \psi = \bar\psi \psi - \barpsi$. 
In the thermodynamic limit, it takes different values depending on the type of phase transition. $B_4(\bar\psi \psi)=1$ corresponds to a first-order phase transition,  $B_4(\bar\psi \psi)=3$ to an analytic crossover, and $B_4(\bar\psi \psi)= 1.604$ to the second order phase transition with 3-dimensional Z(2) universality class~\cite
{Blote:1995zik}.
However, at finite volume, $B_4(\bar{\psi} \psi)$ exhibits volume dependence for first-order transitions and crossover. It approaches the corresponding universal value in the thermodynamic limit. At the critical point of a second-order phase transition, the value remains unchanged with volume, it is scale invariant. 

\begin{figure}[!htp]
\centering
\includegraphics[width=0.38\textwidth]{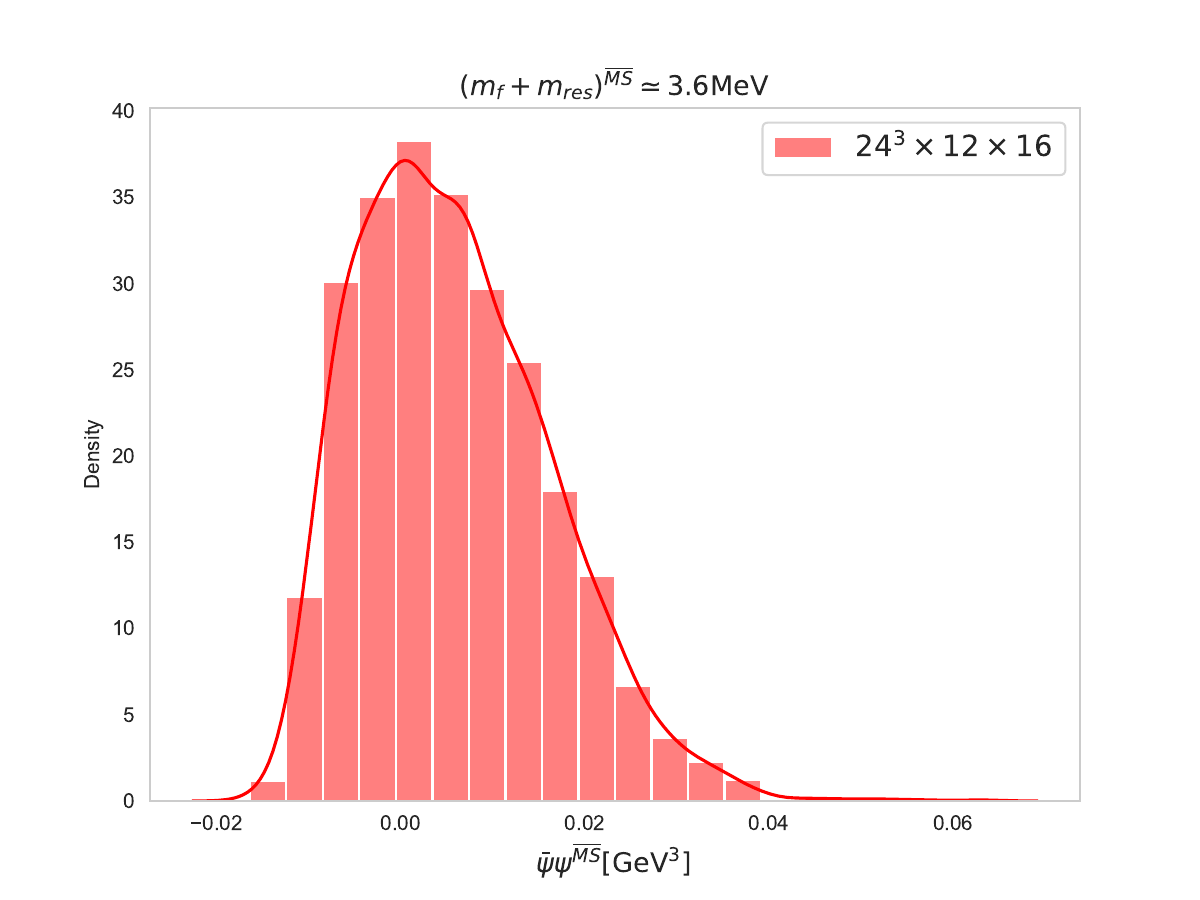}~
\includegraphics[width=0.38\textwidth]{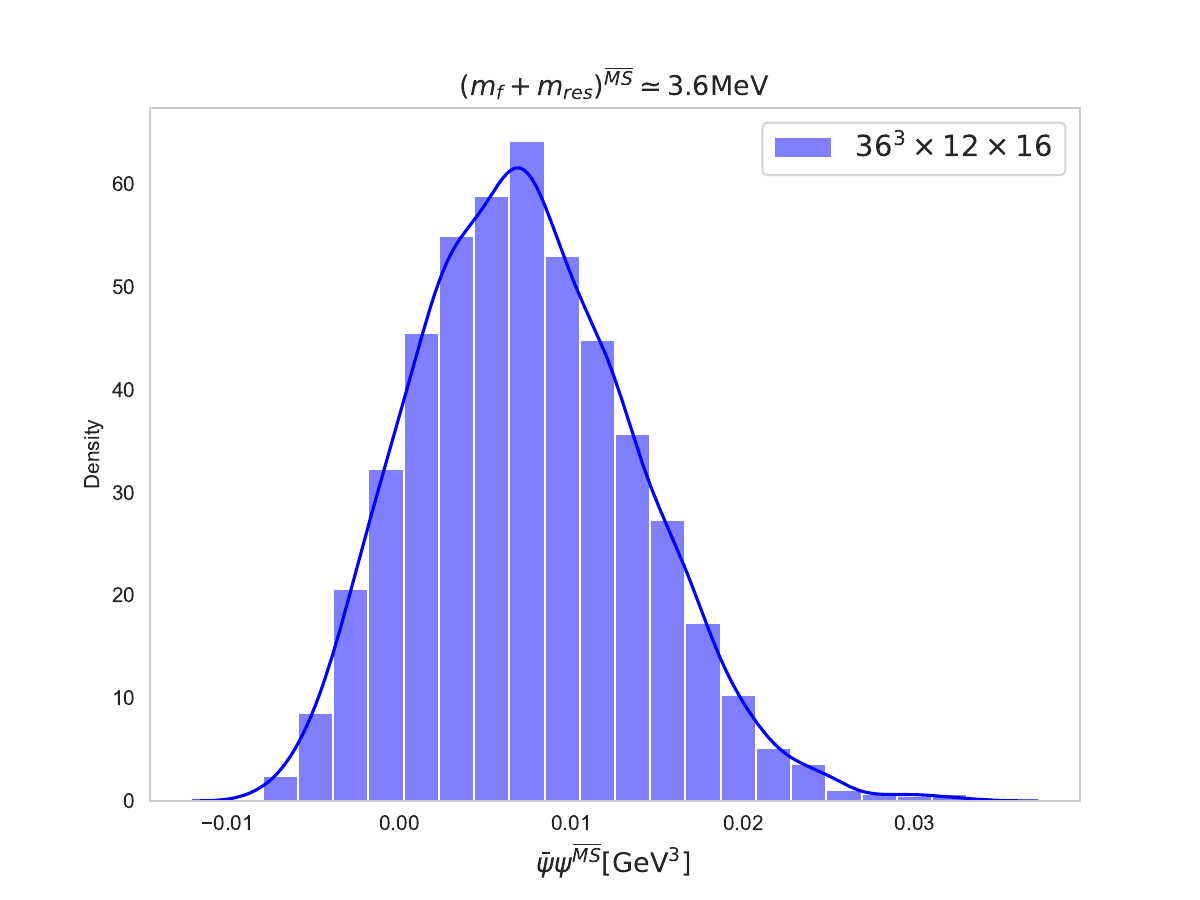}~
\includegraphics[width=0.38\textwidth]{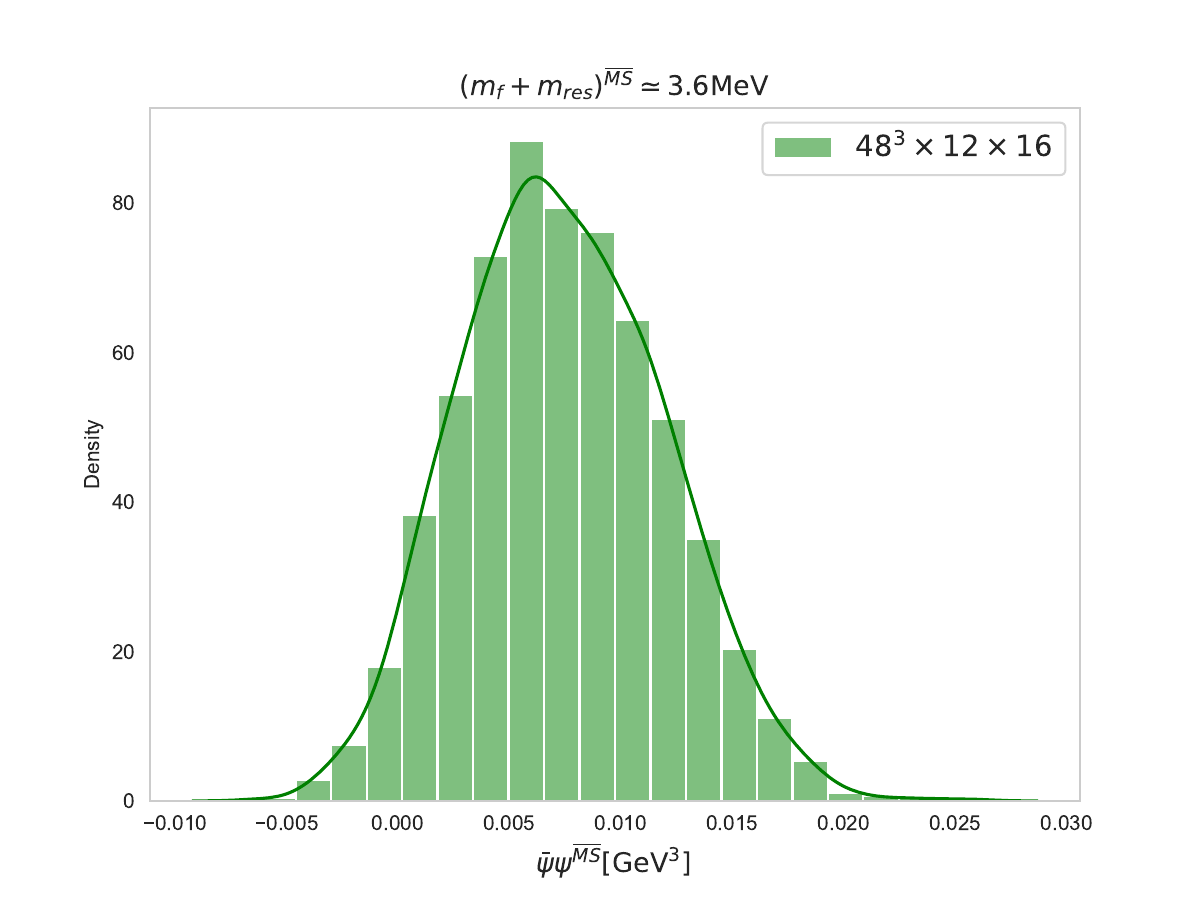}
\caption{The histogram of chiral condensate in the vicinity of the transition mass point  $(m_f + \mres)^{\overline{\mathrm {MS}}}(2\,\mathrm{GeV}) \sim 3.6$ MeV for $N_t=12$ lattices with three different volumes.}
\label{fig:histogram}
\end{figure}

The right panel of~\autoref{fig:chidisc} shows $B_4(\bar{\psi} \psi)$ as a function of quark mass for $N_t=12$ lattices. Results from all the ensembles are close to around the transition mass point, which is $(m_f + \mres)^{\overline{\mathrm {MS}}}(2\,\mathrm{GeV}) \sim 3.6$ MeV as determined from the peak of disconnected chiral susceptibility, indicating a crossover transition. The order of the QCD phase transition can also be studied using the histogram the chiral condensate at the transition point. \autoref{fig:histogram} displays the distribution of chiral condensate near the transition point 3.6 MeV for $N_t=12$ lattices with three different volume. We observe that it behaves like a Gaussian distribution, with no evidence of a double-peak structure, which is expected for a first-order phase transition, would become more pronounced with increasing volume. This provides further evidence for a crossover transition.

\section{Summary}
We studied the $N_f=3$ QCD phase transition using M\"{o}bius domain wall fermions at a fixed lattice spacing of $a=0.1361(20)$ on $48^3\times 12\times 16$ lattices with a range of quark masses. Through the analysis of the chiral condensate, disconnected chiral susceptibility, and Binder cumulant, and by comparing these results with those obtained on $24^3 \times 12 \times 16$ and $36^3\times 12\times 16$ lattices in our previous works~\cite{Zhang:2022kzb,Zhang:2024ldl}, we 
observe that the transtion is consistent with a smooth crossover. The inflection 
point is located at approximately $m_f^{\mathrm{\overline {MS}}}(2\, \mathrm{GeV}) \sim 4$ MeV, at a temperature of 121(2) MeV. This result is consistent with findings obtained using Wilson and staggered-type fermions. 


To investigate residual chiral symmetry breaking effects, we performed additional simulations on $24^3\times 12\times 32$ lattices at $\beta=4.0$. We observed that the $1/L_s$ dependence dominates the contribution to the residual mass at this strong coupling. 
By examining the residual chiral symmetry breaking effects on chiral observables for 
$L_s=16$ and 32 lattices, we found that the disconnected chiral susceptibility results are consistent for approximately the same total quark mass across both $L_s=16$ and 32. In contrast, the chiral condensate results are larger for $L_s=32$ than for 16, indicating reduced residual chiral symmetry breaking for larger $L_s$. This behavior is expected, as the disconnected chiral susceptibility does not suffer from UV divergence, while the chiral condensate does. By obtaining the coefficients ($C^D$, $x$) of the UV divergence term in the chiral condensate, expressed as $C^D \frac{m_f + x\mres}{a^2}$, 
we explicitly subtracted this unwanted UV divergence term from the domain wall fermion chiral condensate for the first time. This is demonstrated by the consistent renormalized chiral condensate results at the same total quark mass between the $L_s=16$ and $L_s=32$ lattices.



\acknowledgments
We would like to thank our collaborators in the JLQCD collaboration for their helpful discussions. This research benefitted from the computational resources made available by several institutions and projects. These include the Supercomputer Fugaku, provided by the RIKEN Center for Computational Science through the HPCI projects hp220233 and hp210032, as well as Usability Research ra000001. The work also utilized the Wisteria/BDEC-01 Supercomputer System at Tokyo University/JCAHPC through HPCI project hp220108, the Ito supercomputer at Kyushu University through HPCI projects hp190124 and hp200050, and the Hokusai BigWaterfall at RIKEN. This work was supported in part by JSPS KAKENHI grant Nos. 20H01907 and 21H01085. Additionally, author Yu Zhang acknowledges support from the Deutsche Forschungsgemeinschaft (DFG, German Research Foundation) through the CRC-TR 211 "Strong-interaction matter under extreme conditions" – project number 315477589 – TRR 211. We also acknowledge the Grid Lattice QCD framework\footnote[1]{https://github.com/paboyle/Grid} and its extension tailored for A64FX processors~\cite{Meyer:2019gbz}, which played a crucial role in generating the QCD configurations for this research. We thank N. Meyer and T. Wettig for their valuable discussions regarding the use of Grid for A64FX. For the measurement, we utilized the 
Hadrons code~\cite{antonin_portelli_2023_8023716}, 
which significantly contributed to the success of this work.

\bibliographystyle{JHEP.bst}
\bibliography{ref.bib}

\providecommand{\href}[2]{#2}\begingroup\raggedright\begin{thebibliography}{10}

\bibitem{Zhang:2022kzb}
Y.~Zhang, Y.~Aoki, S.~Hashimoto, I.~Kanamori, T.~Kaneko and Y.~Nakamura,
  \emph{{Finite temperature QCD phase transition with 3 flavors of Mobius
  domain wall fermions}}, \href{https://doi.org/10.22323/1.430.0197}{\emph{PoS}
  {\bfseries LATTICE2022} (2023) 197}
  [\href{https://arxiv.org/abs/2212.10021}{{\ttfamily 2212.10021}}].

\bibitem{Zhang:2024ldl}
Y.~Zhang, Y.~Aoki, S.~Hashimoto, I.~Kanamori, T.~Kaneko and Y.~Nakamura,
  \emph{{Exploring the QCD phase diagram with three flavors of M\"obius domain
  wall fermions}}, \href{https://doi.org/10.22323/1.453.0203}{\emph{PoS}
  {\bfseries LATTICE2023} (2024) 203}
  [\href{https://arxiv.org/abs/2401.05066}{{\ttfamily 2401.05066}}].

\bibitem{Pisarski:1983ms}
R.D.~Pisarski and F.~Wilczek, \emph{{Remarks on the Chiral Phase Transition in
  Chromodynamics}}, \href{https://doi.org/10.1103/PhysRevD.29.338}{\emph{Phys.
  Rev. D} {\bfseries 29} (1984) 338}.

\bibitem{Fejos:2022mso}
G.~Fejos, \emph{{Second-order chiral phase transition in three-flavor quantum
  chromodynamics?}},
  \href{https://doi.org/10.1103/PhysRevD.105.L071506}{\emph{Phys. Rev. D}
  {\bfseries 105} (2022) L071506}
  [\href{https://arxiv.org/abs/2201.07909}{{\ttfamily 2201.07909}}].

\bibitem{Fejos:2024bgl}
G.~Fejos and T.~Hatsuda, \emph{{Order of the SU(Nf)\texttimes{}SU(Nf) chiral
  transition via the functional renormalization group}},
  \href{https://doi.org/10.1103/PhysRevD.110.016021}{\emph{Phys. Rev. D}
  {\bfseries 110} (2024) 016021}
  [\href{https://arxiv.org/abs/2404.00554}{{\ttfamily 2404.00554}}].

\bibitem{Kousvos:2022ewl}
S.R.~Kousvos and A.~Stergiou, \emph{{CFTs with $\mathbf{U(m)\times U(n)}$
  global symmetry in 3D and the chiral phase transition of QCD}},
  \href{https://doi.org/10.21468/SciPostPhys.15.2.075}{\emph{SciPost Phys.}
  {\bfseries 15} (2023) 075}
  [\href{https://arxiv.org/abs/2209.02837}{{\ttfamily 2209.02837}}].

\bibitem{Pisarski:2024esv}
R.D.~Pisarski and F.~Rennecke, \emph{{Conjectures about the Chiral Phase
  Transition in QCD from Anomalous Multi-Instanton Interactions}},
  \href{https://doi.org/10.1103/PhysRevLett.132.251903}{\emph{Phys. Rev. Lett.}
  {\bfseries 132} (2024) 251903}
  [\href{https://arxiv.org/abs/2401.06130}{{\ttfamily 2401.06130}}].

\bibitem{Giacosa:2024orp}
F.~Giacosa, G.~Kov\'acs, P.~Kov\'acs, R.D.~Pisarski and F.~Rennecke,
  \emph{{Anomalous U(1)A couplings and the Columbia plot}},
  \href{https://doi.org/10.1103/PhysRevD.111.016014}{\emph{Phys. Rev. D}
  {\bfseries 111} (2025) 016014}
  [\href{https://arxiv.org/abs/2410.08185}{{\ttfamily 2410.08185}}].

\bibitem{Bernhardt:2023hpr}
J.~Bernhardt and C.S.~Fischer, \emph{{QCD phase transitions in the light quark
  chiral limit}},
  \href{https://doi.org/10.1103/PhysRevD.108.114018}{\emph{Phys. Rev. D}
  {\bfseries 108} (2023) 114018}
  [\href{https://arxiv.org/abs/2309.06737}{{\ttfamily 2309.06737}}].

\bibitem{Liao:2001en}
X.~Liao, \emph{{Study of 3 flavor QCD finite temperature phase transition with
  staggered fermions}},
  \href{https://doi.org/10.1016/S0920-5632(01)01735-2}{\emph{Nucl. Phys. Proc.
  Suppl.} {\bfseries 106} (2002) 426}
  [\href{https://arxiv.org/abs/hep-lat/0111013}{{\ttfamily hep-lat/0111013}}].

\bibitem{Karsch:2001nf}
F.~Karsch, E.~Laermann and C.~Schmidt, \emph{{The Chiral critical point in
  three-flavor QCD}},
  \href{https://doi.org/10.1016/S0370-2693(01)01114-5}{\emph{Phys. Lett. B}
  {\bfseries 520} (2001) 41}
  [\href{https://arxiv.org/abs/hep-lat/0107020}{{\ttfamily hep-lat/0107020}}].

\bibitem{deForcrand:2003vyj}
P.~de~Forcrand and O.~Philipsen, \emph{{The QCD phase diagram for three
  degenerate flavors and small baryon density}},
  \href{https://doi.org/10.1016/j.nuclphysb.2003.09.005}{\emph{Nucl. Phys. B}
  {\bfseries 673} (2003) 170}
  [\href{https://arxiv.org/abs/hep-lat/0307020}{{\ttfamily hep-lat/0307020}}].

\bibitem{Iwasaki:1995ij}
Y.~Iwasaki, K.~Kanaya, S.~Sakai and T.~Yoshie, \emph{{Chiral phase transition
  in lattice QCD with Wilson quarks}},
  \href{https://doi.org/10.1007/s002880050179}{\emph{Z. Phys. C} {\bfseries 71}
  (1996) 337} [\href{https://arxiv.org/abs/hep-lat/9504019}{{\ttfamily
  hep-lat/9504019}}].

\bibitem{Jin:2014hea}
X.-Y.~Jin, Y.~Kuramashi, Y.~Nakamura, S.~Takeda and A.~Ukawa, \emph{{Critical
  endpoint of the finite temperature phase transition for three flavor QCD}},
  \href{https://doi.org/10.1103/PhysRevD.91.014508}{\emph{Phys. Rev. D}
  {\bfseries 91} (2015) 014508}
  [\href{https://arxiv.org/abs/1411.7461}{{\ttfamily 1411.7461}}].

\bibitem{Karsch:2003va}
F.~Karsch, C.R.~Allton, S.~Ejiri, S.J.~Hands, O.~Kaczmarek, E.~Laermann et~al.,
  \emph{{Where is the chiral critical point in three flavor QCD?}},
  \href{https://doi.org/10.1016/S0920-5632(03)02659-8}{\emph{Nucl. Phys. B
  Proc. Suppl.} {\bfseries 129} (2004) 614}
  [\href{https://arxiv.org/abs/hep-lat/0309116}{{\ttfamily hep-lat/0309116}}].

\bibitem{deForcrand:2006pv}
P.~de~Forcrand and O.~Philipsen, \emph{{The Chiral critical line of N(f) = 2+1
  QCD at zero and non-zero baryon density}},
  \href{https://doi.org/10.1088/1126-6708/2007/01/077}{\emph{JHEP} {\bfseries
  01} (2007) 077} [\href{https://arxiv.org/abs/hep-lat/0607017}{{\ttfamily
  hep-lat/0607017}}].

\bibitem{Varnhorst:2015lea}
L.~Varnhorst, \emph{{The $N_f$=3 critical endpoint with smeared staggered
  quarks}}, \href{https://doi.org/10.22323/1.214.0193}{\emph{PoS} {\bfseries
  LATTICE2014} (2015) 193}.

\bibitem{Bazavov:2017xul}
A.~Bazavov, H.T.~Ding, P.~Hegde, F.~Karsch, E.~Laermann, S.~Mukherjee et~al.,
  \emph{{Chiral phase structure of three flavor QCD at vanishing baryon number
  density}}, \href{https://doi.org/10.1103/PhysRevD.95.074505}{\emph{Phys. Rev.
  D} {\bfseries 95} (2017) 074505}
  [\href{https://arxiv.org/abs/1701.03548}{{\ttfamily 1701.03548}}].

\bibitem{Jin:2017jjp}
X.-Y.~Jin, Y.~Kuramashi, Y.~Nakamura, S.~Takeda and A.~Ukawa, \emph{{Critical
  point phase transition for finite temperature 3-flavor QCD with
  non-perturbatively O($a$) improved Wilson fermions at $N_{\rm t}=10$}},
  \href{https://doi.org/10.1103/PhysRevD.96.034523}{\emph{Phys. Rev. D}
  {\bfseries 96} (2017) 034523}
  [\href{https://arxiv.org/abs/1706.01178}{{\ttfamily 1706.01178}}].

\bibitem{Kuramashi:2020meg}
Y.~Kuramashi, Y.~Nakamura, H.~Ohno and S.~Takeda, \emph{{Nature of the phase
  transition for finite temperature $N_{\rm f}=3$ QCD with nonperturbatively
  O($a$) improved Wilson fermions at $N_{\rm t}=12$}},
  \href{https://doi.org/10.1103/PhysRevD.101.054509}{\emph{Phys. Rev. D}
  {\bfseries 101} (2020) 054509}
  [\href{https://arxiv.org/abs/2001.04398}{{\ttfamily 2001.04398}}].

\bibitem{Cuteri:2021ikv}
F.~Cuteri, O.~Philipsen and A.~Sciarra, \emph{{On the order of the QCD chiral
  phase transition for different numbers of quark flavours}},
  \href{https://doi.org/10.1007/JHEP11(2021)141}{\emph{JHEP} {\bfseries 11}
  (2021) 141} [\href{https://arxiv.org/abs/2107.12739}{{\ttfamily
  2107.12739}}].

\bibitem{Dini:2021hug}
L.~Dini, P.~Hegde, F.~Karsch, A.~Lahiri, C.~Schmidt and S.~Sharma,
  \emph{{Chiral phase transition in three-flavor QCD from lattice QCD}},
  \href{https://doi.org/10.1103/PhysRevD.105.034510}{\emph{Phys. Rev. D}
  {\bfseries 105} (2022) 034510}
  [\href{https://arxiv.org/abs/2111.12599}{{\ttfamily 2111.12599}}].

\bibitem{Nakamura:2022abk}
Y.~Nakamura, Y.~Aoki, S.~Hashimoto, I.~Kanamori, T.~Kaneko and Y.~Zhang,
  \emph{{Finite temperature phase transition for three flavor QCD with
  M\"obius-domain wall fermions}},
  \href{https://doi.org/10.22323/1.396.0080}{\emph{PoS} {\bfseries LATTICE2021}
  (2022) 080}.

\bibitem{Meyer:2019gbz}
N.~Meyer, D.~Pleiter, S.~Solbrig and T.~Wettig, \emph{{Lattice QCD on upcoming
  Arm architectures}}, \href{https://doi.org/10.22323/1.334.0316}{\emph{PoS}
  {\bfseries LATTICE2018} (2019) 316}
  [\href{https://arxiv.org/abs/1904.03927}{{\ttfamily 1904.03927}}].

\bibitem{Borsanyi:2012zs}
S.~Borsanyi et~al., \emph{{High-precision scale setting in lattice QCD}},
  \href{https://doi.org/10.1007/JHEP09(2012)010}{\emph{JHEP} {\bfseries 09}
  (2012) 010} [\href{https://arxiv.org/abs/1203.4469}{{\ttfamily 1203.4469}}].

\bibitem{Furman:1994ky}
V.~Furman and Y.~Shamir, \emph{{Axial symmetries in lattice QCD with Kaplan
  fermions}}, \href{https://doi.org/10.1016/0550-3213(95)00031-M}{\emph{Nucl.
  Phys. B} {\bfseries 439} (1995) 54}
  [\href{https://arxiv.org/abs/hep-lat/9405004}{{\ttfamily hep-lat/9405004}}].

\bibitem{Blum:2000kn}
T.~Blum et~al., \emph{{Quenched lattice QCD with domain wall fermions and the
  chiral limit}}, \href{https://doi.org/10.1103/PhysRevD.69.074502}{\emph{Phys.
  Rev. D} {\bfseries 69} (2004) 074502}
  [\href{https://arxiv.org/abs/hep-lat/0007038}{{\ttfamily hep-lat/0007038}}].

\bibitem{RBC:2008cmd}
{\scshape RBC, UKQCD} collaboration, \emph{{Localization and chiral symmetry in
  three flavor domain wall QCD}},
  \href{https://doi.org/10.1103/PhysRevD.77.014509}{\emph{Phys. Rev. D}
  {\bfseries 77} (2008) 014509}
  [\href{https://arxiv.org/abs/0705.2340}{{\ttfamily 0705.2340}}].

\bibitem{Sharpe:2007}
S.R.~Sharpe, \emph{Future of chiral extrapolations with domain wall fermions},
  \href{https://arxiv.org/abs/0706.0218}{{\ttfamily 0706.0218}}.

\bibitem{Binder:1981sa}
K.~Binder, \emph{{Finite size scaling analysis of Ising model block
  distribution functions}}, \href{https://doi.org/10.1007/BF01293604}{\emph{Z.
  Phys. B} {\bfseries 43} (1981) 119}.

\bibitem{Blote:1995zik}
H.W.J.~Blote, E.~Luijten and J.R.~Heringa, \emph{{Ising universality in three
  dimensions: a Monte Carlo study}},
  \href{https://doi.org/10.1088/0305-4470/28/22/007}{\emph{J. Phys. A}
  {\bfseries 28} (1995) 6289}
  [\href{https://arxiv.org/abs/cond-mat/9509016}{{\ttfamily
  cond-mat/9509016}}].

\bibitem{antonin_portelli_2023_8023716}
A.~Portelli, N.~Lachini, felixerben, mmphys, F.~Joswig, rrhodgson et~al.,
  \emph{aportelli/hadrons: Hadrons v1.4},  June, 2023.
\newblock 10.5281/zenodo.8023716.

\end{thebibliography}\endgroup



\end{document}